\documentclass[english]{article}
\usepackage[latin9]{inputenc}
\usepackage{geometry}
\usepackage{array}
\usepackage{float}
\usepackage{textcomp}
\usepackage{amssymb}
\usepackage{graphicx}
\usepackage{setspace}
\makeatletter

\providecommand{\tabularnewline}{\\}

\makeatother

\usepackage{babel}
\begin{document}

\title{Wage gap between men and women in Tunisia}

\author{Hela Jeddi and Dhafer Malouche\footnote{Ecole Sup\'erieure de la Statistique et de l'Analyse de l'Information, Unit\'e Mod\'elisation et Analyse Statistique et Economique, University of Carthage, Tunisia}}
\maketitle
\begin{abstract}
This paper focuses on estimating wage differences between males and
females in Tunisia by using the Oaxaca-Blinder decomposition, a technical
that isolates wage gap due to characteristics, from wage gap due to
discrimination against women. The data used in the analysis is obtained
from the Tunisian Population and Employment Survey 2005. It is estimated
that, the gender wage gap is about 19\% and the results ascertain
that the gender wage gap is mostly attributed to discrimination, especially
to underestimation of females'caracteristics on the labor market.
\end{abstract}

\paragraph*{Keywords: }

Gender wage inequality, Blinder-Oaxaca decomposition technical, linear
regression models, R statistical programming language

\section{Introduction}

Tunisia has always been considered as one of the most advanced Arab
countries in terms of women rights, especially owing to its family
code enacted in 1956. Indeed, right after independence, the social
dimension was indispensable to the creation of a modern and balanced
society which prohibits all forms of exclusion and ensures equal opportunities,
including the improvement of demographic and health indicators, lengthening
life expectancy, the increase in immunization coverage, the decline
in infant mortality and reducing poverty and expanding coverage. The
coutnry then ratified the Convention on the Elimination of All Forms
of Discrimination against Women in 1985 and amended the labor code,
the penal code, nationality code, etc. which hav strengthened the
rights of women in Tunisia. The new Tunisian Constitution of January
2014 seems to be continuing the same tradition and is showing advances
in the field women's rights, especially with two articles (21 and
46) which decide on discrimination, equal opportunities in the positions
of responsibility and gender-based violence.

If the adoption of the Personal Status Code in 1956, which took place
well before the establishment of the republic, formalized gender equality,
the impact of the ensuing education system performance improvement
was spectacular. Women\textquoteright s education has altered their
course and social relations in general. Tunisia has undergone major
socioeconomic transformations which have upset the traditional patriarchal
structure, impacting the status of women in the family and in society.
As a result of this State Feminism, Tunisian women benefit from higher
status and greater opportunities in education and employment. However,
in 2011, with the revolution and the political unrest that followed,
we discover a very different reality and perceive a profound social
malaise that can easily play in favor of religiosity and probably
to the detriment of women. After more than half a century of feminism
and political and civil commitments, the sitution seems to be weak
and could then toggle to the disadvantage of women. So one may wonder
whether changes in women's progress in demographic transition and
access to education and maternal health resources have effectively
resulted in the integration of women into economic and political activities.
This situation can be analyzed from the perspective of the labor market.
Girls\textquoteright{} vocational streams are increasingly diversified,
as they have more opportunities to access specialities of their choice,
even those that are traditionnaly dominated by men. However, the change
has not reflected in the labor market which largely remains influenced
by professional male and female stereotypes and continues to offer
females training courses and professions perceived as being appropriate
to their gender (education, health, social service, etc.), In this
market, women are threatened by job insecurity, unemployment, and
frequently find themselves in a vicious circle: their chances to find
work are relatively small, they become discouraged, and sometimes
give up by leaving the labor market when it\textquoteright s difficult
to balance work and family life. How can we account for the poor economic
integration of women into the labor market? How can we measure the
amount of discrimination against them? Why is gender discrimination
such is issue, when, since independence, the rights given to women
were supposed to be part of a wider policy of development and modernization
of the country?

The paper is divided in five sections. The first section introduces
the problem of gender wage discrimination in Tunisia. Section II describes
the Oaxaca-Blinder decomposition. In section III, we briefly review
the data used for the wage equation estimation and the methodology
employed. Section IV concerns the empirical model. Finally section
V presents the results and the relevant conclusions.

1- Gender wage discrimination in Tunisia.

Questions on the status of women are not new in Tunisia and women's
rights restoration was the first measure of independent Tunisia. Indeed,
the promulgation of the Personal Status Code on 13 August 1965 upset
many widely held Tunisian beliefs on several levels (abolition of
polygamy, institution of legal divorce, removing the right to impose
marriage, fixing the age minimum for marriage at 18 years for girls,
subject to his consent and custody given to the mother if the father
dies, institution of adoption and guardianship ...). This code was
the first important law enacted after independence, even before the
proclamation of the republican regime in July 1957.

Consequently, the legislative reforms in women's rights and gender
equality were particularly important in the history of Tunisia after
independence. Indeed, assuming that equal participation in all spheres
of life is an essential element for development, Bourguiba explained
that any objective, which is based on democracy, respect for human
rights and sustainable development could be achieved only if it guarantees
to all women the full enjoyment of their human rights and ensures
that no woman is denied the right to work. These ideas were certainly
not new for Tunisia, but Bourguiba had the desire to make reforms.
Indeed, in 1857, Mohamed Senoussi wrote a study on women in Islam
to discuss women's rights in Muslim religion. The author focused on
and was committed to the education of women, despite the requirements
at that time which prohibited them from writing prose or poetry. He
shows that education of women is important, not only for the fulfillment
of their religious duties, but also for their obligation to bring
up children and help their husband. One can also mention Tahar Hadad,
who continued this reformist movement. In 1930, 30 years before the
promulgation of CSP, he presented his bold and modernist ideas and
thus began a courageous fight for the legal and social liberation
of women. Tahar Haddad assessed a divided and inert society that lives
on the memory of a sublimated past prohibiting any development based
on time requirements (see \cite{haddad1930}). He criticized traditional
Muslim laws such as polygamy, repudiation and forced marriage, which
had, according to Haddad, deteriorated the status of Muslim women.
In addition, he encouraged a new reading of sacred texts away from
a simple exegetical and literalist reading while advocating for the
adoption of a new analysis that would search for the goals and higher
objectives of Sharia. Between 1980 and 2011, while barricading freedom
and weakening the opposition, Ben Ali voluntarily continued the process
initiated by his predecessor in women emancipation. A will which he
demonstrated through the constitutional amendments of 1993, the creation
of a strong institutional framework and the inclusion of the gender
approach to the agenda of the five-year development plans since 1991.

Ben Ali was constantly monitored by feminist movements and Tunisian
civil society, both of which have been vigilant in questioning the
status of women. Paradoxically, the second president, while establishing
an authoritarian and autocratic regime, continued the state feminism
of his predecessor and promoted the empowerment of women. He nevertheless
exploited the theme of women in connection with radical Islam and
terrorism to legitimize attacks on freedom of expression.

On a legal level, it should be noted that Tunisia joined CEDAW on
1985, with reservations, which were repealed by Decree Act in August
2011. The new Tunisian Constitution of January 2014 shows positive
signs, especially with two Articles (21 and 46) which focus on discrimination,
equal opportunities in différent senior positions, and gender-based
violence. 

We therefore find that Tunisian feminist tradition is not new has
been a constant goal long before the revolution. But the big question
remains: Will one of the most equal codes of personal status in the
region actually save women from social and economic vulnerability?
Let\textquoteright s begin with women\textquoteright s situation in
the labor market, where the low status could be approached by wage
differentials between the sexes. The situation will be analyzed through
three characteristics: Female participation, employment and unemployment.

\section{Tunisian women employment caracteristics}

Two measures have played a major role in improving the status of women
in Tunisia and opened the doors to a girl\textquoteright s salaried
work: education and demograaphic policy. Education is one of the fundamental
achievements of the societal project in Tunisia, as young people\textquoteright s
easy access to education has greatly helped the profound social changes
affecting the country since the independence. Massive efforts to enroll
students have been made, notably through the introduction of compulsory
education up to 16 years and equal access to all to education without
discrimination of any kind. Today, the enrollment rates of young children
stand at nearly identical levels for girls (99\%) and boys (98.9\%).
In higher education, the enrollment rate of girls even exceeds that
of boys. Demographic policy, mainly family planning and birth control,
has often been presented as a requirement for improving the status
of women. Thus, in 40 years, the fertility rate fell from 7\% to the
level of 2\%, avoiding an uncontrollable population explosion. The
natural population growth rate decreased from 1.96\% in 1990 to 1.09\%
in 2014 and life expectancy keeps lengthening for both sexes. The
structure of the population by age tends toward strengthening active
age groups (15-59 years) and an accelerated shrinking of the child
population (under 15 years). But apparently, despite increasing enrollment
rates and birth control, inequalities between men and women persist,
especially in the labor market, where progress often lags. Tunisian
women remain pretty inactive and the rate of women's access to the
labor market is relatively slow given the demographic, educational
and legal change made in the country.

\subsection{Persistence of low activity rate.}

On moving from 5.5\% in 1966 to 25.7\% (against 70.3\% for males and
around 47\% for both sexes) in 2012, the participation of women over
15 years old in the labor market has grown, particularly for young
people, but it is too low compared to men\textquoteright s and to
the global average. The progress in training-education has certainly
made women more qualified, but it has not had a major effect on parity.

Except for the boom times it experienced in the 70s with the promulgation
of a law on export industries, women\textquoteright s activity has
largely been limited to unskilled, low-wage jobs. Once again, the
gap between the sexes is in favor of men (table 1). The largest proportion
of women employed in 2011 is that of employees, leading with nearly
79.5\% (68.6\% of employed men are employees), followed by independent
workers at 12.5\% (28.2\% men occupied) and family workers at 8\%
(3.2\% of employed men). 

Even if the gap is getting narrower, it still persists at levels above
those of developed countries. Women\textquoteright s entry into the
labor market resulted in an initial explosion of unemployment and
they mostly concentrated on occupations that erect, in one way or
another, femininity as a professional quality. Thus, with 80\%, salaried
employee is the professional category which hosts the most women.
On the contrary, men beat them in self-employed category. 

\begin{table}[H]
\begin{centering}
\begin{tabular}{|c|c|c|c|c|c|c|c|c|}
\hline 
Year & 1966 & 1975 & 1988 & 1994 & 2004 & 2010 & 2011 & 2012\tabularnewline
\hline 
\hline 
Male & 83.50\% & 81.10\% & 78.60\% & 73.80\% & 67.80\% & 69.50\% & 70.60\% & 70.3\%\tabularnewline
\hline 
Female & 5.50\% & 18.90\% & 21.80\% & 22.90\% & 24.20\% & 24.80\% & 23.70\% & 25.70\%\tabularnewline
\hline 
Whole & 44.90\% & 50.20\% & 50.50\% & 48.40\% & 45.80\% & 46.90\% & 47.80\% & 47.70\%\tabularnewline
\hline 
\end{tabular}
\par\end{centering}

\caption{Evolution of the activity rate by gender ( 15 years old and more),
Source: National employment survey. Institut National des Statistiques
(INS).\label{tab1}}
\end{table}

\subsection{Too few diversified employments }

The increasing participation of women in the labor market can be illustrated
through the employment rate, whose low levels reflect female sex integration
difficulties. Indeed, once again, the rate is well below the developed
countries ones. While male employment is diversified across many sectors,
2/3 of employed women are concentrated in three main sectors with
high female labor, particularly services (49.4\%), manufacturing (26.4\%)
and agriculture (16.7\%); women doing these activities tend to suffer
more climatic and economic setbacks. Indeed, girls career choices
are increasingly diversified and they possess opportunities to access
the specialty of their choice and perform tasks tradionally dominated
by men. However, the change has not been reflected in the labor market,
which is largely influenced by professional male and female stereotypes.
For their part, girls continue to favor training courses and professions
perceived as appropriate to their gender, such as education, health
and social service. In these areas, women think they will find better
employment opportunities and have the ability to reconcile family
and working life. 

As a results, girls often find themselves employed in areas with the
least job security: In 2013, 20\% are in technical sciences compared
to 72.9\% in lettres. In higher education, in a set of 22 sectors,
14 are particularly feminized: social and human sciences, languages,
economics, journalism, law, agriculture, services. Girls are relatively
less represented in engineering (29\%), architect (34.5\%), veterinary
emergency (35.9\%), physical sciences (45.9\%), information technology
(47.4\%), and mathematics and statistics (49.4\%) . Job creation also
benefits men more than women and, during the last five years, women
have only received 17\% of all new jobs, a third of the average job
demand. Even though school enrollment level is the same for boys and
girls, or even better for the latter, differences exist in the curricular
path followed by each sex. The female segment of the labor market
remains poorly diversified and follows a certain sex ratio that limits
women's access to many resources the market provides. Thus, girls
are concentrated in unskilled labor sectors, which not only fosters
increased competition between women but also results in a devaluation
of these sectors.

\subsection{High unemployment rates}

The increasing participation of women, although it remains very low,
has increased pressure on the labor market. The situation was completely
different in the middle 80s when men were more likely to be unemployed
than women (13.7\% against 11\%). Tunisian labor market is essentially
men\textquoteright s market and unemployment affects women more than
men as shown in table (2). In deed, not only the unemployment rate
reduction in recent years has benefited men more than women but also,
the gap in overall unemployment rates between both sexes has been
growing ever since. Women unemployment decreased seriously between
1966 and 1980, owing to the enactment of Law 72 on the establishment
of export industries occupying many of the female population. It then
recorded a slight decrease from19.6\% to 16.2\% between 1999 and 2003,
and resumed its growth in 2003. Today, with a value of 26.9\% against
a global average of 6.5\%, it is among the highest in the world. 

Additionnally to its high level, unemployment particularly affects
graduates of both sexes and it comes as no surprise that women are
the first affected. According to the latest data (the fourth quarter
of 2013) of the National Institute of Statistics (INS), in 2012, unemployed
universities graduates has reached 187.500, whose two-thirds are young
women (67\%). Female university graduates have to wait more than a
year, sometimes even two or three years to be offered a job. At the
same period, among young women (25-34 years) only 41\% are in the
labor market, against 89\% of men in the same age group. In 2013,
the unemployment is more severe among women (21.9\%) than men (12.8\%),
and affects twice as many women university graduates (41.9\%) than
male graduates (21.7\%) These figures, however, remain underestimations
and well below the reality of unemployment among Tunisian women, as
they are not present on the formal labor market. In 2011, 66.3\% of
young women are declared inactive, but only 18\% of unskilled women
are estimated to be unemployed. Half of young women graduates, are
putting much more than a year to find a first job against 32\% of
their male counterparts. Finally, important pay disparities between
men and women persist. There are few studies to identify these differences
accurately, the only one, performed on the data of 1999 population
employment survey by the INS on a sample of 5976 employees including
1441 women highlights a gender pay gap of 18\%. It\textquoteright s
owed to discrimination for a total of 12\%, giving a relative contribution
of 67.65\%. With relative contribution of 32.35\%, productive characteristics
explain nearly 6\% of the gap . 

\begin{table}[H]
\begin{centering}
\begin{tabular}{|c|c|c|c|c|c|c|c|c|}
\hline 
Year & 2005 & 2006 & 2007 & 2008 & 2009 & 2010 & 2011 & 2012\tabularnewline
\hline 
\hline 
Male & 12.10\% & 11.50\% & 11.30\% & 11.20\% & 11.30\% & 10.90\% & 15.00\% & 14.90\%\tabularnewline
\hline 
Female & 15.20\% & 15.10\% & 15.90\% & 15.90\% & 18.80\% & 18.90\% & 27.40\% & 26.60\%\tabularnewline
\hline 
Whole & 12.90\% & 12.50\% & 12.40\% & 12.40\% & 13.30\% & 13.00\% & 18.30\% & 18.10\%\tabularnewline
\hline 
\end{tabular}
\par\end{centering}

\caption{Evolution of the unemployment rate by gender Source: National employment
survey. Institut National des Statistiques (INS)\label{tab2}}
\end{table}

\section{The Oaxaca-Blinder decomposition}

In this section, we review the basics of Oaxaca-Blinder decompositions
and discuss the alternative choices of counterfactual wage structure.
Firstly, we specify that in economics, there is discrimination when
two equally qualified and, thus perfect substitute for each other
individuals are treated differently solely on the basis of a non-economic
characteristic such as their gender, race, ethnicity, disability,
etc. Discrimination acknowledgement as a development of the standard
representation of the labor market, was followed in the 70s by empirical
and experimental methods of measuring discrimination by sex, race
or ethnic origins. The usual procedure is based on the decomposition
technique Oaxaca-Blinder, the work coming out of two economists Alan
Blinder (see \cite{blinder73}) and Ronald Oaxaca (see \cite{oaxaca73}),
who introduced it in 1973 in the economic literature. The work eventually
became the basic tool to study the various forms of discrimination.
The idea is based on the residual difference methodology where workers
with identical characteristics are equally productive and therefore
should expect an identical wage. 

The residual difference method decomposes within the average pay gap
between two populations, one part expressed by differences in productive
characteristics of individuals: the endowments effect, and a second
unjustified part related to differences in identical characteristics
returns on the labor market between both populations, the residual
difference. This unjustified gap is then assimilated to wage discrimination.
The authors estimate separate reduced Mincerian wage equations for
the two compared groups of workers. Both of the groups are supposed
to have the same wage formation process as written in (\ref{eq:OB})

\begin{equation}
\left.\begin{array}{l}
\log W_{m}=X_{m}\alpha_{m}+\epsilon_{m}\\
\log W_{f}=X_{f}\alpha_{f}+\epsilon_{f}
\end{array}\right.\label{eq:OB}
\end{equation}

Each of the equations in (\ref{eq:OB}) is an expression of the linear
regression models that explains the logarithme of the wage of each
category (Female and Male). Indeed $W_{m}$ and $W_{f}$ represent
respectively the vector of wages of the Males and Females in our Sample.
The matrices $X_{m}$ and $X_{f}$ contain respectively the independent
variables of the Male and Female Samples. In both of the them the
first column contains only ones, i.e; $\mathbf{1}=(1,1,\ldots,1)',$
and this column represents the intercept in the regression models
expressed in (\ref{eq:OB}). The vectors $\epsilon_{m}$ and $\epsilon_{f}$
are the errors in the models expressend in (\ref{eq:OB}).They are
supposed to be random vectors satisfying the following assumptions:
zero mean, i.e $\mathbb{E}\left(\epsilon_{m}\right)=\mathbb{E}\left(\epsilon_{f}\right)=0$
and each of them has a diagonal variance matrix where all the diagonal
elements are equal to an unknown parameter $\sigma^{2}>0$. Finally
the parameters $\alpha_{m}$ and $\alpha_{f}$ are the unknown vectors
that will be estimated using the Ordinary Least Squares method (OLS).
Their estimators are expressed as follows

\begin{equation}
\begin{array}{rcl}
\widehat{\alpha}_{m} & = & \left(X_{m}^{'}X_{m}\right)^{-1}X_{m}^{'}\left(\log W_{m}\right)\\
\widehat{\alpha}_{f} & = & \left(X_{f}^{'}X_{f}\right)^{-1}X_{f}^{'}\left(\log W_{f}\right)
\end{array}\label{eq:OB2}
\end{equation}

The estimators computed in (\ref{eq:OB2}) will help us to understand
how the $\log$ of the Wage vary in terms of the characteristics of
the sample. We can then write the difference between the mean of the
log of the wages as a sum of two parts. A first part can be explained
by the difference of the characteristics of the two samples Male and
Females. The second part is the unexplained part of this decomposition.
Indeed we can express the difference between the mean of the log of
the wages as below 

\begin{equation}
\overline{\log W_{m}}-\overline{\log W_{f}}=\widehat{\alpha}_{m}^{'}\overline{X}_{m}-\widehat{\alpha}_{f}^{'}\overline{X}_{f}=\widehat{\alpha}_{m}^{'}\overline{X}_{m}-\widehat{\alpha}_{f}^{'}\overline{X}_{m}+\widehat{\alpha}_{f}^{'}\overline{X}_{m}-\widehat{\alpha}_{f}^{'}\overline{X}_{f}=\underbrace{\left(\widehat{\alpha}_{m}-\widehat{\alpha}_{f}\right)^{'}\overline{X}_{m}}_{\mbox{unexplained part}}+\underbrace{\widehat{\alpha}_{f}^{'}\left(\overline{X}_{m}-\overline{X}_{f}\right)}_{\mbox{explained part}}\label{eq:OB3}
\end{equation}
 where $\overline{X}_{m}$ and $\overline{X}_{f}$ are the vector
of the means of the independant variables of the models in (\ref{eq:OB}).
The explained part in (\ref{eq:OB3}) is the shift effect due to group
membership discrimination in the labor market. The unexplained part
is in (\ref{eq:OB3}) is the share attributable to differences in
endowments returns. Thus, if the returns were equal, the wage gap
can be explained entirely by differences in observable characteristics.
If instead characteristics returns were equal, the average income
gap is entirely explained by structural effects which may possibly
be the consequence of other forms of discrimination. 

The decomposition can be sensitive to the used wage structure and
the type of hypothesized discrimination : nepotism tin favour of men
or discrimination against women. In deed, this method raises the question
of the choice of weighting. The challenge is to be able to determine
\textit{a priori} a standard non-discriminatory returns of individual
characteristics and measure relative to this standard the male advantage,
women's disadvantage, and the share resulting from the difference
of characteristics. With wage discrimination hypothesis, men receive
competitive salaries and thus they are paid their marginal productivity,
but women are underpaid (\ref{eq:OB4}). In this case, the rule or
non discriminating wage standard would be that of men. Yet it is also
possible that we were in a situation of nepotism in favor of men,
a situation in which women receive competitive salaries and men are
overpaid. In such a situation, the coefficients from the women\textquoteright s
earnings functions provide an estimate of the nondiscriminatory wage
structure as in the following Equation (\ref{eq:OB3})

\begin{equation}
\overline{\log W_{m}}-\overline{\log W_{f}}=X_{f}\widehat{\alpha}_{f}-X_{m}\widehat{\alpha}_{m}=\widehat{\alpha}_{m}^{'}\overline{X}_{m}-\widehat{\alpha}_{m}^{'}\overline{X}_{f}+\widehat{\alpha}_{m}^{'}\overline{X}_{f}-\widehat{\alpha}_{f}^{'}\overline{X}_{f}=\underbrace{\left(\widehat{\alpha}_{m}-\widehat{\alpha}_{f}\right)^{'}\overline{X}_{f}}_{\mbox{unexplained part}}+\underbrace{\widehat{\alpha}_{m}^{'}\left(\overline{X}_{m}-\overline{X}_{f}\right)}_{\mbox{explained part}}\label{eq:OB4}
\end{equation}

\section{Sample characteristics}

Our results are based on data of the Survey of population and employment
conducted in 2005 by INS, an official statistical agency. The samples
contain waged workers of both sexes who are more than 18 years old.
Male sample is composed of 29,854 individuals and the size of females
sample is 10,482 individuals. For each respondent in the sample, the
database provides information, including age, educational attainment,
employment status, income, industry and occupation of employment,
marital status and the residential location. The average monthlly
earnings are 382 and 323 TD per month for males and females, respectively 

\begin{figure}[H]
\begin{centering}
\includegraphics[scale=0.4]{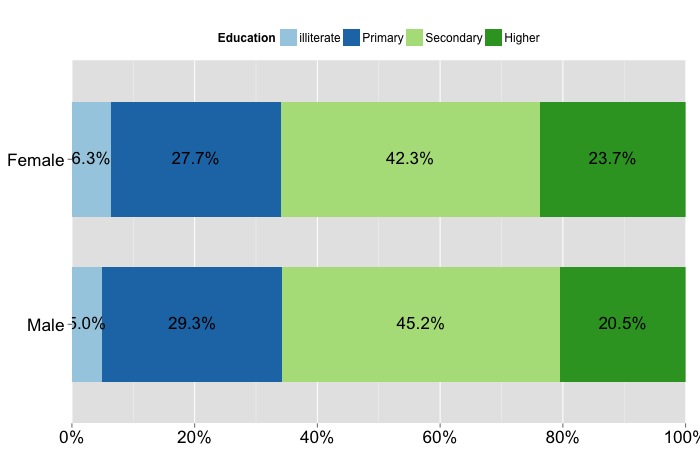}
\par\end{centering}

\caption{\label{fig:Educational-level-by-gender}Educational level for both
gender}
\end{figure}

Men and women are more represented in primary school levels with 64.9\%
and 35.1\% respectively. Among women 20 .7\% are universities graduates
against 12\% for men as shown in Fig \ref{fig:Educational-level-by-gender}

The emergence of new technologies and home appliances has allowed
women to develop their role within the family and has enabled them
to free themselves from household chores. Before, the role of housewife
was always assigned to them and this role takes a lot of time, thus
women had little free time to devote to other business or personal
education. The evolution begins with the advent of more efficient
and therefore less restrictive household tools. Therefore, the working
time of women is reduced. Technological developments is mainly beneficial
to women, because it eases their tasks and allows them to spend the
least possible time. We thus built a new variable that we called the
socieconomic score which provides information on the possession of
equipment and appliance. We expect that the effect is greater on women's
income.

\begin{figure}[H]
\begin{centering}
\includegraphics[scale=0.4]{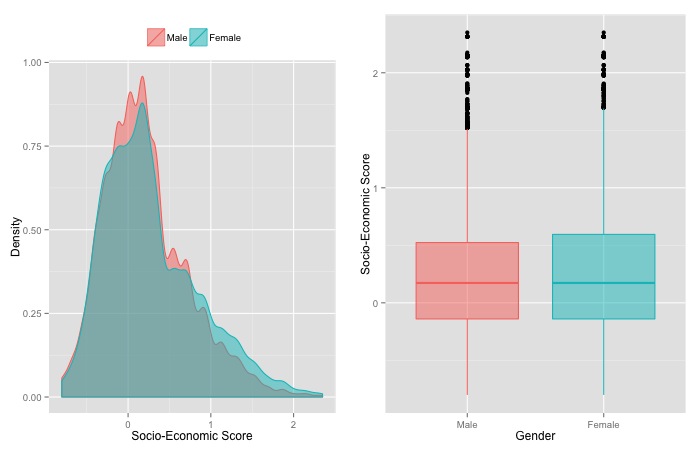}
\par\end{centering}

\caption{\label{fig:Distribution-of-ses by gender}Distribution of the Socio-Economic
Score for both gender}

\end{figure}

We considered the different levels of access to equipments acording
to gender. We took into account ressurces such as cars, TV, washers,
camputers, cookers, fridges, dishwashers, mobiles, access to internet
\dots{} (more details in appendix). Men and women have different access
and control differently the resources depending on their gender, age,
\textit{Social-Economic} group, educational level, ect. Woman\textquoteright s
access to resources is supposed to be delayed and more difficult.
From all these variables we have a Multiple Correspondance Analysis
to construct an Index that measures a \textit{Socio-Economic situation
}of the respondant (see \cite{Husson2010}). These scores take values
in $\mathbb{R}$ where high positive values indicate a high level
of \textit{Socio-Economic situation} where live the respondant and
negatives values indicate low level for the \textit{Social-Economic}
situtation of the respondant. In Figure \ref{fig:Distribution-of-ses by gender}
we make a brief comparison between gender regarding the \textit{Social-Economic}
score. We can easily note that there's no difference between Males
and Females in our Sample. 

\begin{figure}[H]
\begin{centering}
\includegraphics[scale=0.4]{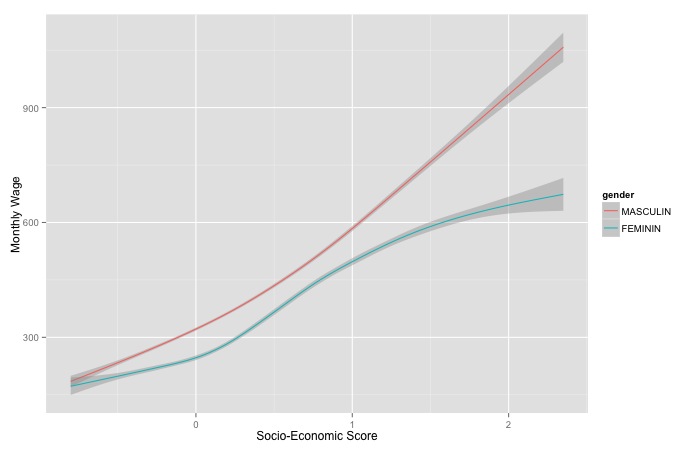}
\par\end{centering}

\caption{\label{fig:Socioeconomic-Score-versus Wage}Socioeconomic Score versus
Monthly wage}

\end{figure}

In Figure \ref{fig:Socioeconomic-Score-versus Wage} we represent
the estimated non linear model used to exaplain the Monthly Wage according
to the \textit{Social-Economic} Score for both gender. We easily notuce
that men salary increases gradually with the \textit{Social-Economic}
score. However for women, not only the effect on wage is less important
but the relatioship is decreasing because of the concavity between
levels of \textit{Social-Economic} score and Monthly remuneration:
a better \textit{Social-Economic} level is worth less and less for
women.

\begin{figure}[H]
\begin{centering}
\begin{tabular}{cc}
\includegraphics[scale=0.3]{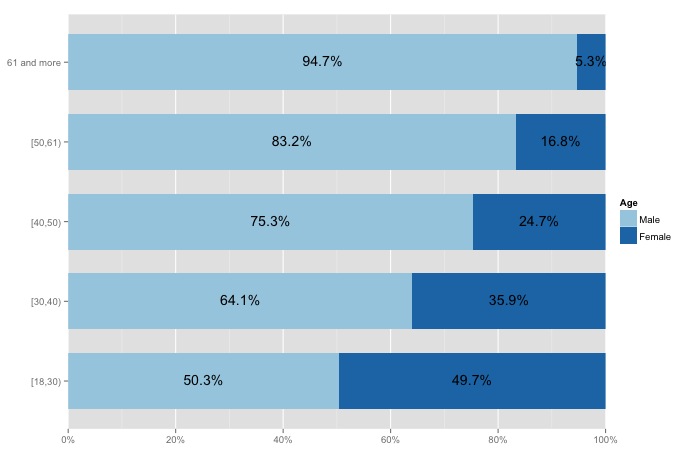} & \includegraphics[scale=0.3]{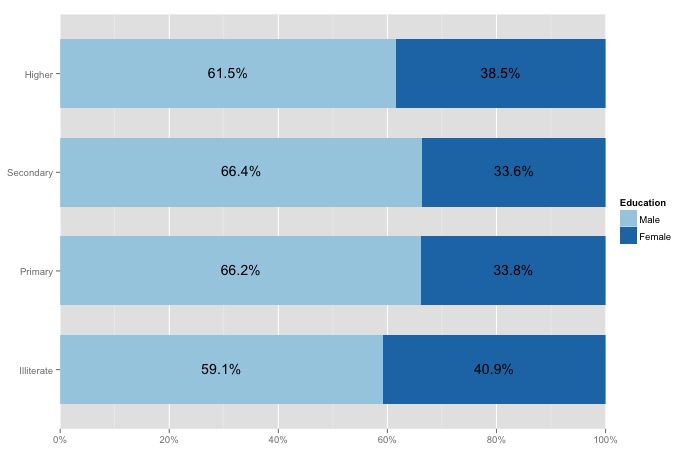}\tabularnewline
\includegraphics[scale=0.3]{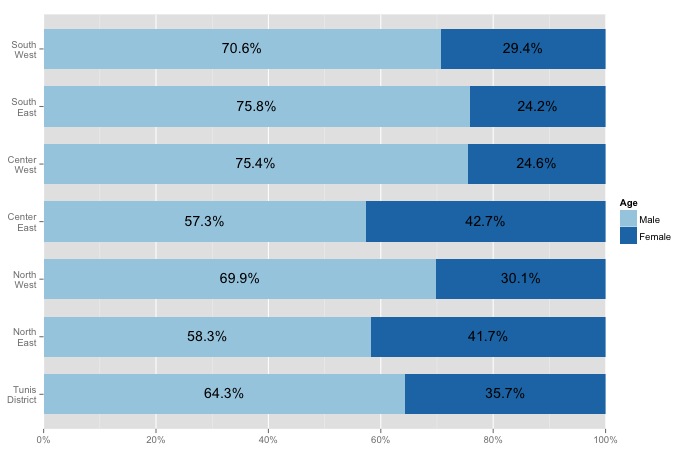} & \includegraphics[scale=0.3]{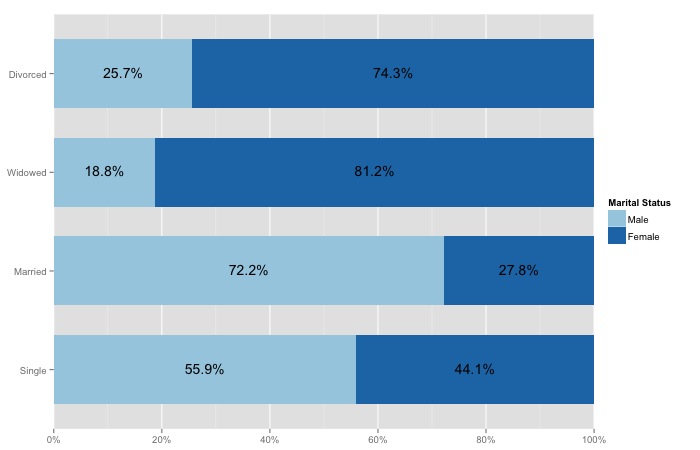}\tabularnewline
\end{tabular}
\par\end{centering}

\caption{\label{fig:Statistics}Structure of the employed population by sex
and by socio-economic characteristics }
\end{figure}

Summary statistics for the used variables  are detayled in Figure
\ref{fig:Statistics}.  

\begin{figure}[H]
\begin{centering}
\includegraphics[scale=0.4]{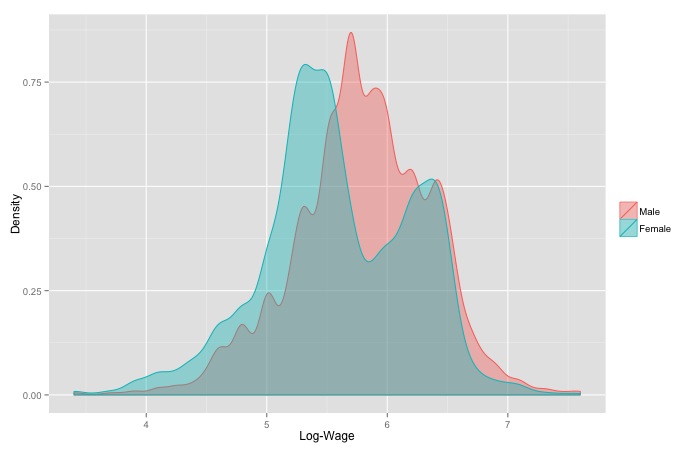}
\par\end{centering}

\caption{\label{fig:The-estimation-of-distibution by Gender}The estimation
of the distribution of the logarithm of the Monthly Wage for both
gender}

\end{figure}

In Figure \ref{fig:The-estimation-of-distibution by Gender} we show
the estimation of the distribution of the logarithme of the monthly
wages for both gender. We first notice that the \textit{Male curve}
is shifted to the left according to the \textit{Female curve}. This
means that the monthly wages of males are overall higher than monthly
wage of females. However by observing the \textit{Female curve} we
can notice two modes (pikes). We can then deduce that the Female sample
could be devided in two homogenious groups. A first one has a lower
monthly wage than the whole Male sample. The second group of females
has a monthly wage almost equal to the Man monthly wages. 

\section{The empirical model}

In Oaxaca-Blinder decomposition, whatever is the non-discriminatory
wage structure used, we have one reference group, one for each component:
the difference in the endowments taking dominant group, thus men,
as the reference group (explained part in equation (\ref{eq:OB3}))
and the difference in the way market value the two groups characteristics,
with low-paid, thus women, as the reference group (unexplained part
in equation (\ref{eq:OB3})). 

The reference group may influence the explained and unexplained parts,
to overcome this problem some writers on the discrimination litterature
re-arange the the mean log-wage equations to transform the simple
bipartite decomposition into a tripartite one: the \textit{three-fold
decomposition} decomposes the difference into three categories difference
: endowments effect, returns effect and interaction between these
two differences. It uses only one reference groupe for the endowments
and returns components, it is female reference in Equation (\ref{eq:3fold-female})

\begin{equation}
\overline{\log W_{m}}-\overline{\log W_{f}}=\underbrace{\widehat{\alpha}_{f}^{'}\left(\overline{X}_{m}-\overline{X}_{f}\right)}_{\mbox{endowments effect}}+\underbrace{\underbrace{\left(\widehat{\alpha}_{m}-\widehat{\alpha}_{f}\right)^{'}\overline{X}_{f}}_{\mbox{coefficient effects}}+\underbrace{\left(\widehat{\alpha}_{m}-\widehat{\alpha}_{f}\right)^{'}\left(\overline{X}_{m}-\overline{X}_{f}\right)}_{\mbox{interaction effects}}}_{\mbox{discrimination}}\label{eq:3fold-female}
\end{equation}

and male reference in the third one as in Equation (\ref{eq:3forld-male})
(see \cite{jones1984})
\begin{equation}
\overline{\log W_{m}}-\overline{\log W_{f}}=\underbrace{\widehat{\alpha}_{m}^{'}\left(\overline{X}_{m}-\overline{X}_{f}\right)}_{\mbox{endowments effect}}+\underbrace{\underbrace{\left(\widehat{\alpha}_{m}-\widehat{\alpha}_{f}\right)^{'}\overline{X}_{m}}_{\mbox{coefficient effects}}+\underbrace{\left(\widehat{\alpha}_{m}-\widehat{\alpha}_{f}\right)^{'}\left(\overline{X}_{m}-\overline{X}_{f}\right)}_{\mbox{interaction effects}}}_{\mbox{discrimination}}\label{eq:3forld-male}
\end{equation}

As in \cite{hlavac2015} the considered the outcome variable is real
monthly wage, the aim of the decomposition is in this case to explain
the difference in mean wages between women and men by the mean values
of explanatory variables which denotes the endowment effect, the returns
effect and interaction as illustrated in (\ref{eq:3fold-female}).

The outcome variable is explained by the following variables : age,
age squared, socioeconomic score, three educational levels (primary,
secondary and highe)r, six regions of residence (North East, North
West, East Center, West Center South East, South West), ten activities
(mechanical and electrical industry, textile, clothing and leather
industry; cottage industry, trade, transport and telecommunications,
hotel and catering, banking and insurance, housing activities, socio-cultural
services, health administration education services) and the marital
status (marreid, widowed and divorced). In Table \ref{tab2} we present
the results of the estimation of the regression models for both gender.

\begin{table}[!htbp] 
\centering    
\begin{tabular}{p{3cm}lcc}  \\
[-1.8ex]\hline  \hline \\
[-1.8ex]   & \multicolumn{2}{c}{\textit{Dependent variable:}} \\
 \cline{2-3}  \\
[-1.8ex] & \multicolumn{2}{c}{Monthly Salary} \\
  & Female & Male \\ 
\\[-1.8ex] & (1) & (2)\\
 \hline \\[-1.8ex]   Age & 5.971$^{***}$ (4.147, 7.795) & 12.705$^{***}$ (11.371, 14.038) \\ 
  Age squared & $-$0.048$^{***}$ ($-$0.073, $-$0.024) & $-$0.126$^{***}$ ($-$0.142, $-$0.111) \\ \hline \\
   SE score & 83.422$^{***}$ (77.812, 89.031) & 129.704$^{***}$ (124.824, 134.584) \\ \hline \\
   Primary & 39.882$^{***}$ (28.323, 51.440) & 48.462$^{***}$ (37.924, 59.000) \\ 
  Secondary & 76.872$^{***}$ (65.277, 88.467) & 109.216$^{***}$ (98.488, 119.944) \\
   Higher & 261.616$^{***}$ (248.374, 274.859) & 328.675$^{***}$ (316.779, 340.572) \\ \hline \\
   North East & $-$4.573 ($-$12.829, 3.683) & $-$2.539 ($-$9.785, 4.706) \\  
 North West & $-$43.878$^{***}$ ($-$53.743, $-$34.013) & $-$30.518$^{***}$ ($-$38.071, $-$22.966) \\
   East Center & $-$7.526$^{**}$ ($-$14.604, $-$0.449) & $-$2.240 ($-$8.394, 3.914) \\  
 West Center & $-$20.342$^{***}$ ($-$33.270, $-$7.415) & $-$2.456 ($-$11.205, 6.292) \\ 
  South East & $-$38.610$^{***}$ ($-$49.462, $-$27.758) & $-$35.444$^{***}$ ($-$42.842, $-$28.046) \\ 
  South West & $-$57.098$^{***}$ ($-$67.568, $-$46.629) & $-$42.706$^{***}$ ($-$50.504, $-$34.909) \\ \hline \\
   Mechanical and electrical industry & 8.352 ($-$13.745, 30.450) & 3.846 ($-$9.934, 17.625) \\  
 Textile, clothing and leather industry & 10.839 ($-$8.956, 30.634) & $-$3.870 ($-$17.291, 9.551) \\  
 Cottage industry & 0.905 ($-$26.486, 28.295) & $-$2.960 ($-$17.178, 11.258) \\  
 Trade & $-$33.727$^{***}$ ($-$55.871, $-$11.583) & $-$18.676$^{***}$ ($-$30.492, $-$6.861) \\   
Transport and telecommunications & 39.912$^{***}$ (14.779, 65.044) & 45.401$^{***}$ (33.140, 57.663) \\ 
 Tourism and catering & 23.785$^{*}$ ($-$1.168, 48.738) & 20.461$^{***}$ (8.540, 32.382) \\
   Banking and insurance & 96.399$^{***}$ (67.260, 125.538) & 113.424$^{***}$ (94.041, 132.806) \\  
 Housing activities & $-$9.726 ($-$35.189, 15.737) & $-$2.357 ($-$17.352, 12.639) \\ 
  Socio-cultural services & $-$47.011$^{***}$ ($-$68.335, $-$25.687) & $-$25.706$^{***}$ ($-$38.577, $-$12.836) \\ 
  Health, administration and education services & 84.555$^{***}$ (64.280, 104.830) & 24.581$^{***}$ (14.201, 34.962) \\ \hline \\    Married & 31.481$^{***}$ (24.916, 38.045) & 24.933$^{***}$ (18.282, 31.584) \\ 
  Widowed & $-$16.530$^{*}$ ($-$34.262, 1.203) & $-$2.246 ($-$39.153, 34.661) \\   
Divorced & 5.747 ($-$11.500, 22.994) & 4.383 ($-$26.864, 35.629) \\  \hline \\
 Constant & 23.718 ($-$14.387, 61.823) & $-$81.385$^{***}$ ($-$109.321, $-$53.448) \\   \hline \\[-1.8ex]  Observations & 10,482 & 19,372 \\  R$^{2}$ & 0.605 & 0.568 \\  Adjusted R$^{2}$ & 0.604 & 0.567 \\  Residual Std. Error & 129.652 (df = 10456) & 144.671 (df = 19346) \\  F Statistic & 641.128$^{***}$ (df = 25; 10456) & 1,017.129$^{***}$ (df = 25; 19346) \\  \hline  \hline \\[-1.8ex]  \textit{Note:}  & \multicolumn{2}{r}{$^{*}p<0.1$; $^{**}p<0.05$; $^{***}p<0.01$} \\  \end{tabular}\caption{The Estimation of the Regression Models of the Monthly Wage.}\label{tab2}  \end{table} 

Regarding the returns of age which may be accounted for by experience
and seniority, the later influences positively both wages. The effect
is paradoxically much stronger for females than for males as an additional
year increases man mean wage by 5.971 TD and woman\textquoteright s
by 12.705 TD. However, paradoxically this effect is much stronger
for females than for males. As for the quadratic association between
age and wages, our estimates show the usual concave shape that highlights
the effect of an extra year on salary: wage increases with age, but
at a decreasing rate and the effect is stronger for females than for
males. The socioeconomic score favours males and females, the effect
is however greater for the laters. Indeed, a higher socioeconomic
level icreases mean earnings of female individual by 129.704 TD and
mean earnings of male one by 83.422 TD. The results imply that education
favours both females and males since mean wage increase with level
of education for both of them. However, finding of women rate of return
that generally exceeds that for men is supported by many studies in
Côte d'Ivoire with Wim P. M. Vijverberg (see \cite{wim93}), in India
with Geeta Gandhi Kingdon (see \cite{geeta2001}) and in fifty six
countries with George Psacharopoulos (see ). Psacharopoulos who reviewed
the rate of return to education and noted that the rate of return
for women generally exceeds that for men in fifty-six developed and
developing countries. He explained saying the more educated the person
is, the more his human capital is important. Table (2) shows education
tends to favour females more than males, all levels included. For
instance, mean female earnings with tertiary education rise by 328.675
TD relative to one with no education while the mean earnings of male
individual with tertiary education level rise by 261.616 TD relative
to one with no education. The difference is less pronounced when it
comes to primary and secondary levels.

In regards to the activity, the survey provides ten categories. Women
benefit from being employed in typical men\textquoteright s job where
the feminization rate is very week such as mecanical and electrical
industrie, women\textquoteright s job have, hoever, a negative impact
in their mean income. For example, being employed in housing activities
reduces female wages on average by 2.357 TD. Being employed in banking
and insurance is in favor for both of sexes with an increase of 113.424
TD on average for females and 96.399 TD for males.

As for residence results, it was no big surprise to find that working
in the capital could be advantageous in terms of remuneration for
both sexes, exceptionally for women. Estimates reported in table (4)
suggest that employees in Tunis receive, on average, higher salaries
than the ones of internal regions. Results show a very significant
impact of living in urban area (North East and East Center) relative
to rural area for both females and males. 

Regarding marital status, being married is a somewhat important factor
of the wage for both sexes, it however drops more on males. If men
case corroborates with Polachek and Mincer (1974) studies, women\textquoteright s
remains padoxical. Widohood is to no one\textquoteright s advantage,
the effect is however greater on men's income which deacreases by
an average of 16.530TD (against 2.246 TD for women). Divorced men
seem to earn on average more than their divorced female counterparts.

\section{Oaxaca-Blinder gender wage gap decomposition results}

Oaxaca points out the well-known index number problem: It is difficult
to say à priori which of the two wage structures (Equations (\ref{eq:OB3})
and (\ref{eq:OB4})) is the non-discrimination wage structure, since
it is unclear on what basis the salary would be determined if there
was no discrimination. If female wage structure is nondiscriminatory,
we have pure discrimination where males earn more than they should.
On the other hand, male wage structure gives nepotisme in which females
earn less than they should. The latter seems to be more tangible as
we are interested in increasing woman's wage rather than decreasing
man's. In the following, we use the female wage structure as reference
group and assume that all labor market should give anyone with same
qualities women wage equation.

\subsection{Estimation using the female wage structure as reference group}

The results of the estimation of gender wage differentials using the
Oaxaca-Blinder decomposition technique with female wage structure
as reference point, are illustrated in Table \ref{tab:Oaxaca-Blinder-Wage-Decomposition 2fold}.
The mean value of real monthly wages amounted to 381.615 Tunisian
Dinars for males and 323.029 Tunisian Dinars for females in 2005,
the wage gap is therefore about 58.585 Tunisian Dinars. The estimation
of the model implies that, without nepotism, men monthly wages should
equal 335.427 Tunisian Dinars. This means too that, due to nepotism,
females are receiving 46.118 Tunisian Dinars less in their real monthly
wages. The value of discrimination represents 14.3\% of the mean value
of real monthly wages they are actually receiving.

\begin{table}[H]
\begin{centering}
\begin{tabular}{|>{\centering}m{3cm}|c|}
\hline 
Wage decomposition components & \multicolumn{1}{c|}{Males}\tabularnewline
\hline 
\hline 
Mean value of real monthly wages  & 381.615 \tabularnewline
\hline 
\hline 
Overall wage gap & \multicolumn{1}{c|}{58.585}\tabularnewline
\hline 
Due to Characteristics & 12.397 \tabularnewline
\hline 
\hline 
Due to returns on characteristics & 46.518\tabularnewline
\hline 
\hline 
Interaction effect  & -0.330\tabularnewline
\hline 
\hline 
Due to Discrimination & 46.188\tabularnewline
\hline 
\hline 
Wage without discrimination effect & \multicolumn{1}{c|}{335.427}\tabularnewline
\hline 
\end{tabular}
\par\end{centering}

\caption{\label{tab:Oaxaca-Blinder-Wage-Decomposition 2fold}Oaxaca-Blinder
Wage Decomposition Results (Female as the reference group)}

\end{table}

\begin{figure}[H]
\begin{centering}
\includegraphics[scale=0.4]{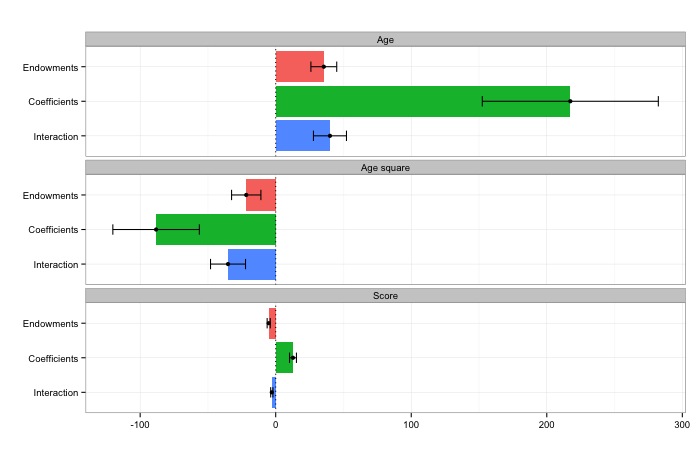}
\par\end{centering}

\caption{\label{fig:Threefold-decomposition-byAge-SE}Threefold decomposition
by Age and Socioeconomic score}
\end{figure}

Next, we examine the endowments and coeffcients components of the
threefold decomposition variable by variable. Figures below show the
estimation results for each variable, along with error bars that indicate
95\% confidence intervals. Age return, assimilited to experience and
seniority, appears to have significant portion of the male-female
wage gap. A less important portion is driven by score returns as shown
in bar graph of Figure \ref{fig:Threefold-decomposition-byAge-SE}.

\begin{figure}[H]
\begin{centering}
\includegraphics[scale=0.4]{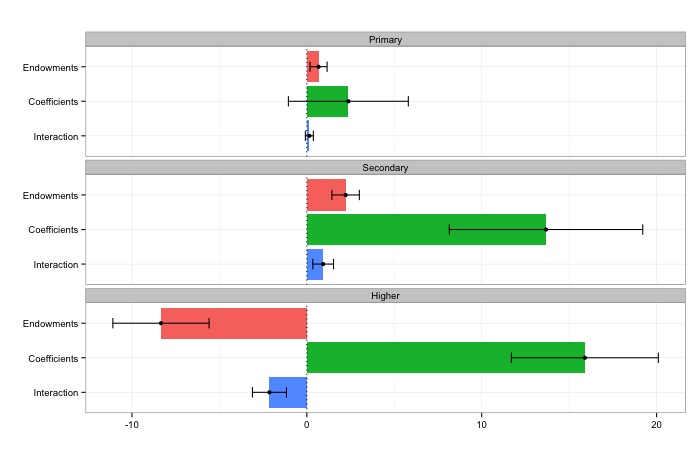}
\par\end{centering}

\caption{\label{fig:Threefold-decomposition-byEducation}Threefold decomposition
by Educational level}

\end{figure}

Regarding educational level as shown in Figure \ref{fig:Threefold-decomposition-byEducation},
in the endowments component, primary and secondary levels appear to
have a statistically significant influence in favour of man, with
the sole exception of tertiary level : women are better endowed than
men, but the market undervalues their characteristics. In the coefficients
component, the three levels achieve clear statistical significance
with most significant portion for tertiary and secondary levels. In
this way, a significant gender wage gap is driven by group differences
in the proportion of individuals with secondary and high school educations.
In other words, individuals with lower diploma tend to earn less.

\begin{figure}[H]
\begin{centering}
\includegraphics[scale=0.4]{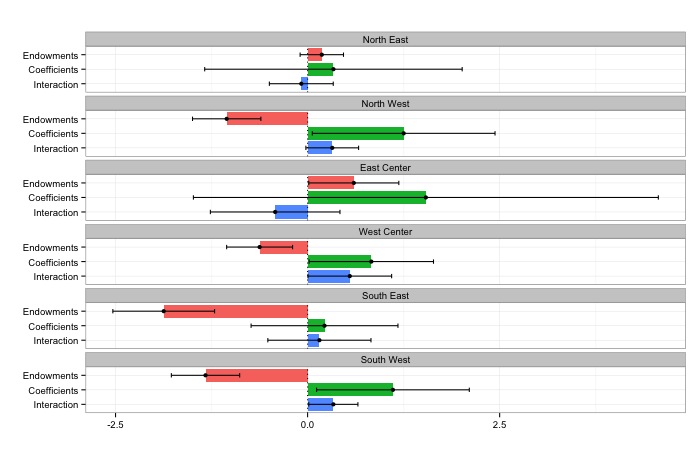}
\par\end{centering}

\caption{\label{fig:Threefold-decomposition-byRegion}Threefold decomposition
by region of residence}

\end{figure}

When we consider region perspective (see Figure \ref{fig:Threefold-decomposition-byRegion}),
the 8th plot graphic shows better endowed women in all regions except
North-East and East-Center. Coefficients effect however harms women
in all regions whatever are their encowments. 

\begin{figure}[H]
\begin{centering}
\includegraphics[scale=0.6]{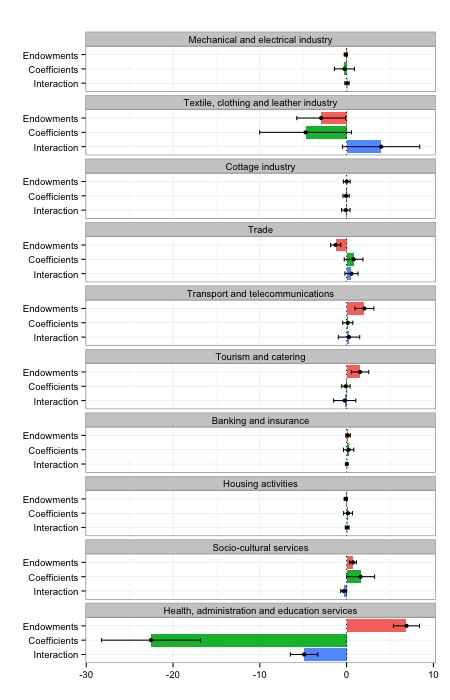}
\par\end{centering}

\caption{\label{fig:Threefold-decomposition-byActivity}Threefold decomposition
by nature of the activity}
\end{figure}

As Figure \ref{fig:Threefold-decomposition-byActivity} makes clear,
differences in the endowments components in transport and telecommunication,
tourism and health, administration and education services account
for the decisive portion of the wage gap explained by endowments.
Results confirm the greater the proportion of women in an occupation,
the lower the characteristics value. The masculine image of an occupation
is a good signal for skills nd professions. Feminized ones are decommissionned.
Effects on coefficients seem to be statistically insignificant or
only marginally significant except for health, administration and
education services, where despite being less endowed than men, market
gives a good estimate to women\textquoteright s characteristics.

The threefold decomposition by marital status in Figure \ref{fig:-Threefold-decompositionStatus}
below shows that a significant portion of the gender wage gap is due
by and large to endowment component of mariage. In the coefficients
component, widowhood effect is the most significant on gap. 

\begin{figure}[H]
\begin{centering}
\includegraphics[scale=0.4]{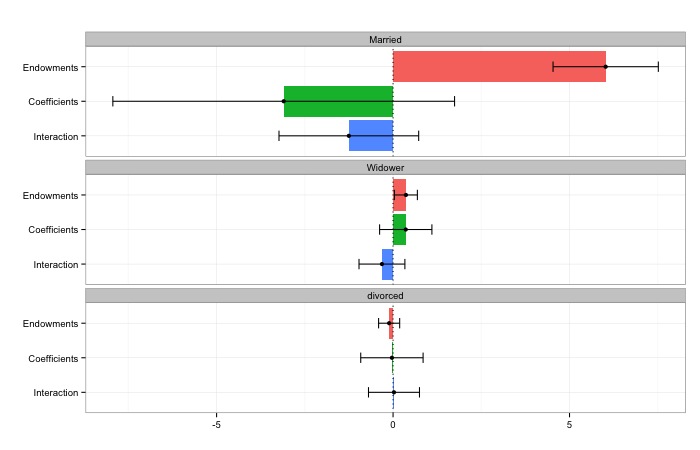}
\par\end{centering}

\caption{\label{fig:-Threefold-decompositionStatus} Threefold decomposition
by marital status}
\end{figure}

Bar graphs (Fig \ref{fig:Educational-level-by-gender}-\ref{fig:-Threefold-decompositionStatus})
detailing characteristic, Bar graphs (9-15) detailing the effects
of characteristics, returns and interaction, could be resumed in the
table below (5). We can sum up by saying that the constant that theorists
interpret as human capital base salary without any human capital endowment
shows an advantage for women in terms of a very important performance
effect (more than the double of discrimination). Thus, at the entrance
of the labor market, pay inequalities are favorable to women. At job
application moment, men and women are supposed to have the same characteristics
but the market valuates better female performance. Experience approximated
by the variable age is an important source of wage inequality with
an effect five times higher than discrimination. It is, however, softened
by the effect of marginal returns: the decreasing marginal returns
of experience attenuate the wage gap between men and women and the
accumulated experience stock reduces the wage gap. Regarding education,
all levels are in favor of men and therefore contribute to wage discrimination.
This discrimination is doubled at higher levels because women have
better characteristics, but the market underestimates their performance.
The regional effect is small but always in favor of men. Marriage
synonymous with family responsibilities and therefore less clamp to
the labor market, is paradoxically favorable to women thanks to its
returns\textquoteright{} effects. 

Married men characteristics are better than women\textquoteright s,
but the trend reverses with coefficients effect. Paradoxically being
married is favourable to women contrary to what has been predicted
by Polachek and Mincer who explained the considerable influence of
the marriage but only on male earnings, and put forward the key role
of career breaks, and maternal penalty in the depreciation of human
capital and deficit of professional experience especially among married
women and mothers. The explanation of such a result could be given
by O.Donni S.Ponthieux (see \emph{\cite{donny2911}}) who wrote that
married woman is dedicated to domestic activities and thus accumulates
a specific capital to couple life. However, because of divorce and
the trend of fluid families, the value of this capital is low and
the spouse might be encouraged to over-invest in the labor market
and accumulate general capital rather than the specific one. Goldin
(see \cite{goldin1990}) gave the same explanation writing that with
divorce rate increasing constantly, women in couples no longer have
a non-market alternative and behave more like single women by anticipatng
a likely separation. Married women try to maximize their bargaining
power by improving their situation in the event of a separation. This
would explain their general willingness to invest in their own careers
and be financially independent and thus to invest in the market work
to the detriment of domestic work. Thus, the university no longer
acts as a marriage market for them. By extending Becker's ideas, Ponthieux
and Donni wrote it would be best that one of the two spouses concerns
himself to domestic activities and thus accumulates a specific capital
to life couple. However, in case of divorce, the value of this capital
is low and this is why both spouses might be encouraged to over-invest
in the labor market and accumulate general capital rather than specific
one.

The wage gap by industry is in favor of women in education, administration
and health services. In these three services although women have smaller
endowments, the market overestimates characteristics compared with
males. I thus recognizes women for qualities they cannot have. The
social construction of female identity in occupations continues and
the market thinks that women do better in women's professional fields
that are associated with feminine qualities. Finally, if women had
the same characteristics as men, on average, their salaries increase
by 12,400 TD, whereas it would increase by 46,500 TD if they received
men remuneration of these characteristics. The Tunisian labor market
pays very poorly the characteristics of women.

\begin{table}[H]
\centering 
\begin{tabular}{|p{3cm}|r|r|r|r|}  
\hline  & characteristic effect & Returns effect & Interaction effect & Nepotism \\  
 \hline (Intercept) & 0.000 & -105.103 & -0.000 & -105.103 \\  
 Age & 35.499 & 217.359 & 40.032 & 257.390 \\   
Age squared & -21.739 & -88.242 & -35.172 & -123.415 \\ 
 SE-score & -5.164 & 12.745 & -2.865 & 9.880 \\
\hline
\multicolumn{5}{|l|}{Education} \\
\hline
Primary & 0.670 & 2.374 & 0.144 & 2.518 \\   
Secondary & 2.219 & 13.676 & 0.934 & 14.609 \\   
Higher & -8.340 & 15.898 & -2.138 & 13.760 \\
 \hline
\multicolumn{5}{|l|}{Region of residence} \\
\hline
North East & 0.185 & 0.336 & -0.082 & 0.254 \\  
 North West & -1.053 & 1.250 & 0.321 & 1.571 \\ 
  East Center & 0.601 & 1.539 & -0.422 & 1.117 \\   
West Center & -0.625 & 0.829 & 0.549 & 1.378 \\    
South East & -1.873 & 0.220 & 0.154 & 0.374 \\ 
 South West & -1.329 & 1.111 & 0.335 & 1.446 \\  
\hline
\multicolumn{5}{|l|}{Nature of the activity} \\
\hline  
Mechanical and electrical industry & -0.087 & -0.246 & 0.047 & -0.199 \\   
Textile, clothing and leather industry & -2.927 & -4.726 & 3.972 & -0.754 \\ 
  Cottage industry & 0.020 & -0.063 & -0.087 & -0.150 \\   
Trade& -1.256 & 0.807 & 0.560 & 1.367 \\   
Transport and telecommunications & 2.051 & 0.136 & 0.282 & 0.418 \\  
 Tourism and catering & 1.545 & -0.082 & -0.216 & -0.298 \\ 
   Banking and insurance & 0.169 & 0.234 & 0.030 & 0.264 \\   
Housing activities & -0.090 & 0.169 & 0.068 & 0.237 \\   
Socio-cultural services & 0.742 & 1.612 & -0.336 & 1.276 \\   
Health, administration and education services & 6.903 & -22.543 & -4.896 & -27.439 \\
\hline
\multicolumn{5}{|l|}{Status} \\
\hline
   Married & 6.025 & -3.098 & -1.253 & -4.352 \\ 
  Widower & 0.364 & 0.360 & -0.315 & 0.045 \\   
Divorced& -0.111 & -0.032 & 0.026 & -0.006 \\  
  \hline 
Overall & 12.397 &	46.518	&-0.330	&46.189 \\
\hline
\end{tabular} 
\caption{\label{tab:Threefold-decomposition-variable}Threefold decomposition
variable by variable (female as the reference group)}

\end{table}

\section{Conclusion }

Tunisia is the most experienced country in terms of respect for women's
rights in the region. The Personal Status Code guaranteed the Tunisian
women all human rights. Parity is gained in terms of access to health
care and schooling. Nevertheless, in the labor market, the situation
is not so idyllic. Women represent 26\% of the work force and the
female unemployment rate is well beyond that of men. Otherwise Tunisian
women earn 16 \% less than male couterparts. It is thus of interest
to know the proportion of these wage differentials that is due to
differences in endowments and the propotion due to discrimination,
that is, the proportion unexplained by caracteristics, and to consider
whether the decomposition is depending on the used non-discriminatory
wage structure. For this purpose, we have introduced the oaxaca package
for the R statistical programming language for linear models to estimate
Blinder-Oaxaca threefold decompositions. We have dealt with differences
in mean wages across two groups. Doing so, we provided estimates for
a detailed, variable-by-variable decomposition and summarized graphically
the results of the decompositions. The conclusion is significant :
gender wage gap is mostly attributed to discrimination and especially
to underestimation of females\textquoteright caracteristics. 

We find the wage gap attributable to returns is more important in
accounting for wage differences than differences attributable to characteristics
in both structures. The wage gap is driven largely by greater returns
to age accounted for by experience and seniority, it\textquoteright s
squared term, by the group differences in secondary and high school
educations and by occupation in health, education and administration
occupations. Women\textquoteright s better position in secondary and
higher educations could contribute to narrowing the gender gap, but
they are underevaluated by the market. Suprisingly, in education,
health and administation services, market , overstimetes women\textquoteright s
caracteristics although they are worse. This shows the impact of the
development level of mentality, especially the perception of the economic
role of women and the keeping of the gendered horizontal segregation
and the gendered roles in professional practices. The marriage variable
results are also interesting in the sense that the effect of the characteristics
confirmed the widespread perception of adverse consequences of marriage
on women's involvement in the productive sphere (traditional division
of roles between the sexes, specializing in housework, accumulation
of specific human capital and weak attachment to the labor market).
Marriage return is however in favor of women. With the help of technology,
increased trends of fluid families and salary levels, the specialization
model is no more effective. Specializing in housework became rather
inefficient and the long term costs of full specialization may outweigh
the short term benefits. In the manner of the wife of the Beckerian
\char`\"{}Treatise on the family\char`\"{}, married woman is no more
dedicated to domestic activities and prefers accumulating general
capital rather than a specific one to guard against the risks of a
probable divorce.

\end{document}